\documentclass[11pt]{article}
\usepackage{moriond,epsfig}

\def\be{\begin{equation}}
\def\ee{\end{equation}}
\def\bea{\begin{eqnarray}}
\def\eea{\end{eqnarray}}

\newcommand{\pmin}{p_{\mbox{\footnotesize{min}}}}

\newcommand{\mujnp}{\mbox{$\mu$-$j$-$\nu$}}

\newcommand{\jjjj}{\mbox{$j$-$j$-$j$-$j$}}

\newcommand{\jjjjnp}{\mbox{$j$-$j$-$j$-$j$-$\nu$}}

\begin{document}
\vspace*{4cm}
\title{MODEL INDEPENDENT SEARCHES IN EP COLLISIONS}

\author{E.~SAUVAN\\
On behalf of the H1 and ZEUS Collaborations}

\address{CPPM, IN2P3-CNRS et Universit\'e de la M\'editerran\'ee,\\
163 Av. de Luminy, F-13288 Marseille, France.}

\maketitle\abstracts{
The high energy program of the HERA collider ended in March $2007$.
In total the H1 and ZEUS experiments collected an integrated luminosity of about $1$ fb$^{-1}$.
Recent results of model independent searches for new physics from both 
experiments are presented. 
Specifically, studies of the events with an isolated lepton and missing transverse momentum
and multi-lepton topologies, where H1 and ZEUS data are combined, and a general signature based search are discussed.
}

\section{Introduction}

At HERA  electrons (or positrons) collide with protons at a centre-of-mass energy of $\sqrt s \simeq 320$~GeV.
During the two running periods of HERA from $1994$ to $2000$ and from $2003$ to $2007$, respectively, the H1 and ZEUS experiments have each recorded $\sim 0.5$ fb$^{-1}$ of data in total, shared between $e^+p$ and $e^-p$ collision modes.
These high energy electron-proton interactions provide a testing ground for the Standard Model (SM) complementary to $e^+e^-$ and $p\bar{p}$ scattering studied at other colliders, giving access to rare processes with cross sections below $1$~pb.
They are therefore used to pursue a rich variety of searches for new phenomena. 
Among them, signature based searches look for differences in precise comparisons between data and SM expectations in different event topologies. 
As an advantage, such model independent analyses do not depend on any a priori definition of expected signatures for exotic phenomena.
Following this approach, final states corresponding to rare SM processes such as real $W$ boson or lepton pair production are investigated.
A general scan at high transverse momenta ($P_T$) of all possible final states is also performed by H1.

\section{Events with high $P_T$ isolated leptons}

The production of $W$ bosons in $ep$ collisions at HERA has a cross-section of about $1$ pb. The leptonic decay of the $W$ leads to events with an isolated high transverse momentum lepton (electron, muon or tau) and missing total transverse momentum. Of particular interest are events with a hadronic system of large transverse momentum ($P_T^X$). An abnormally large rate of high $P_T^X$ events is observed by the H1 experiment~\cite{isollep_h1,H1_prelim_isollep} in the electron and muon channels. In the analysis of all HERA I and HERA II data sets, which amounts to a total luminosity of $478$~pb$^{-1}$, $24$ events are observed at $P_T^X > 25$ GeV for a SM expectation of $15.8 \pm 2.5$. Amongst them only $3$ events are observed in $e^-p$ collisions, in agreement with the SM expectation of $6.9 \pm 1.0$, while $21$ events are observed in the $e^+p$ data for an expectation of $8.9 \pm 1.5$ (see table~\ref{tab:isol_lep_h1zeus}). 

\begin{table*}
\caption{Comparison of the number of isolated lepton (electron or muon) events observed for $P_T^X > 25$ GeV by H1 and ZEUS experiments with SM predictions.}
\begin{center}
\begin{tabular}{|c|ccc|}
\hline
                 & Electron & Muon & Combined\\
$P_T^X > 25$ GeV & obs./exp. & obs./exp. & obs./exp. \\
\hline
~~H1~~ ~$e^-p$~~ $184$ pb$^{-1}$ & $3$ / $3.8 \pm 0.6$ & $0$ / $3.1 \pm 0.5$  & $3$ / $6.9 \pm 1.0$\\
ZEUS ~$e^-p$~~ $206$ pb$^{-1}$ & $3$ / $3.2 \pm 0.6$ & $2$ / $2.4 \pm 0.4$   & $6$ / $5.6 \pm 1.0$\\
\hline
~~H1~~ ~$e^+p$~~ $294$ pb$^{-1}$ & $11$ / $4.7 \pm 0.9$ & $10$ / $4.2 \pm 0.7$ & $21$ / $8.9 \pm 1.5$\\
ZEUS ~$e^+p$~~ $286$ pb$^{-1}$ & $3$ / $3.9 \pm 0.6$ & $3$ / $3.6 \pm 0.5$     & $6$ / $7.5 \pm 1.1$\\
\hline
\end{tabular}
\end{center}
\label{tab:isol_lep_h1zeus}
\end{table*}

The ZEUS experiment has carried out a similar analysis using $492$~pb$^{-1}$ of $1996$--$2007$ data~\cite{ZEUS_isollep}. 
The results are also shown in table~\ref{tab:isol_lep_h1zeus}. At $P_T^X > 25$~GeV the number of data events observed by ZEUS is in agreement with the SM expectation in both $e^+p$ and $e^-p$ collisions. 
A detailed comparison between efficiencies of the H1 and ZEUS detectors for the $W$ signal was performed. Both efficiencies are comparable in the central region. While H1 detection region extends to lower polar angle than ZEUS, most of the high $P_T^X$ events observed by H1 are within the range of the ZEUS acceptance.

\begin{figure}[htbp] 
  \begin{center}
    \includegraphics[width=.45\textwidth]{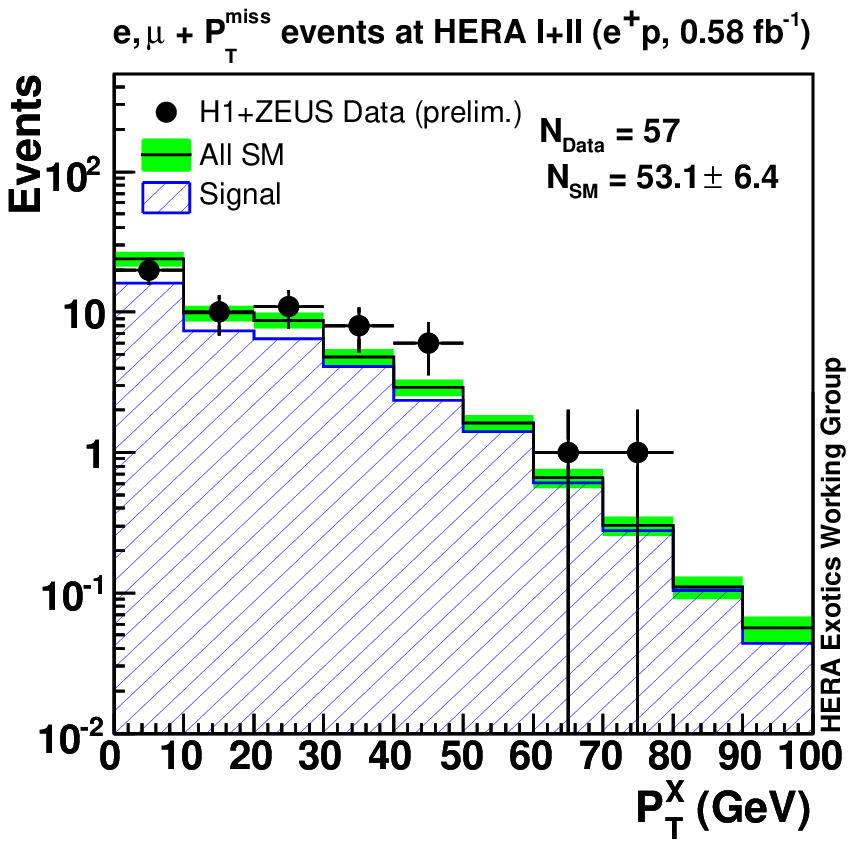}\put(-40,130){{\bf (a)}}
    \includegraphics[width=.45\textwidth]{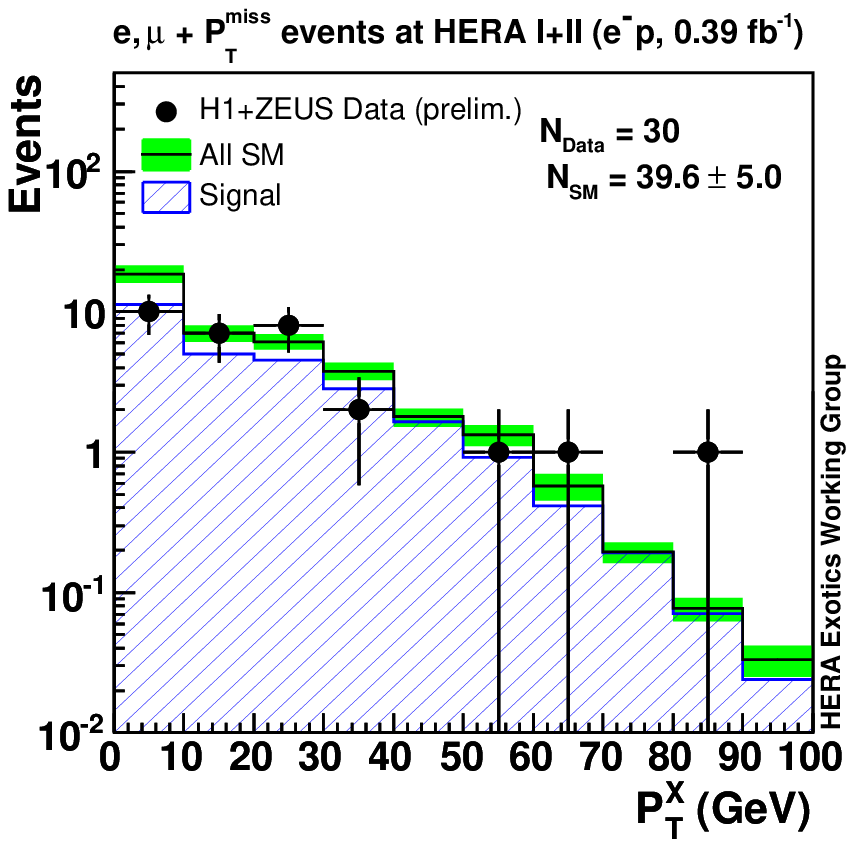}\put(-40,130){{\bf (b)}}
  \end{center}
  \caption{Hadronic transverse momentum distribution of isolated lepton events observed by H1 and ZEUS in $e^+p$ (a) and $e^-p$ (b) data samples. The total SM expectation is represented by the open histograms and the contribution from $W$ production by the hatched histogram. }
\label{fig:isollep}  
\end{figure}

The data samples of the H1 and ZEUS experiments have been used for a combined analysis performed in a common phase space~\cite{H1ZEUS_isollep}. The combined data set corresponds to a total integrated luminosity of $0.97$~fb$^{-1}$.
A total of $87$ events containing an isolated electron or muon and missing transverse momentum are observed in the data, compared to a SM expectation of $92.7 \pm 11.2$. At  $P_T^X > 25$~GeV, a total of $29$ events are observed compared to a SM prediction of $25.3 \pm 3.2$. In this kinematic region, $23$ events are observed in the $e^+p$ data compared to a SM prediction of $14.6 \pm 1.9$. 
The observations in the $e^+p$ and $e^-p$ data sets are exemplified in figure~\ref{fig:isollep} where the $P_T^X$ distributions of both data sets are displayed.

The analysis of the tau decay channel is also performed by H1~\cite{H1_prelim_tau} on all HERA data with a total luminosity of $471$~pb$^{-1}$. In this channel, the separation of the $W$ signal from other SM processes is more difficult and  the purity and efficiency are lower than for the $e$ and $\mu$ channels. In total, $20$ data events are observed compared to a SM expectation of $19.5 \pm 3.2$. One of the data events has $P_T^X$ above $25$ GeV, compared to a SM expectation of $0.99 \pm 0.13$.
An older analysis of the tau channel performed by the ZEUS Collaboration~\cite{Chekanov:2003bf} on HERA~I data reported an observation of two data events with $P_T^X > 25$~GeV, compared to a SM expectation of $0.2 \pm 0.05$.

\section{Multi-lepton events}

The main production mechanism for multi-lepton events is photon-photon collisions. All event topologies with high transverse momentum electrons and muons have been investigated by the H1 experiment~\cite{H1_mlep} using a total luminosity of $459$ pb$^{-1}$. 
The measured yields of di-lepton and tri-lepton events are in good agreement with the SM prediction, except in the tail of the distribution of the scalar sum of transverse momenta  of the leptons ($\sum P_T$). In $e^+p$ collisions, $4$  data events with at least two high $P_T$ leptons are observed with $\sum P_T > 100$ GeV compared to a SM prediction of $1.2 \pm 0.2$. No such events are observed in $e^-p$ collisions for a similar SM expectation of $0.8 \pm 0.2$.

\begin{figure}[htbp] 
\begin{center}
\includegraphics[width=.7\textwidth]{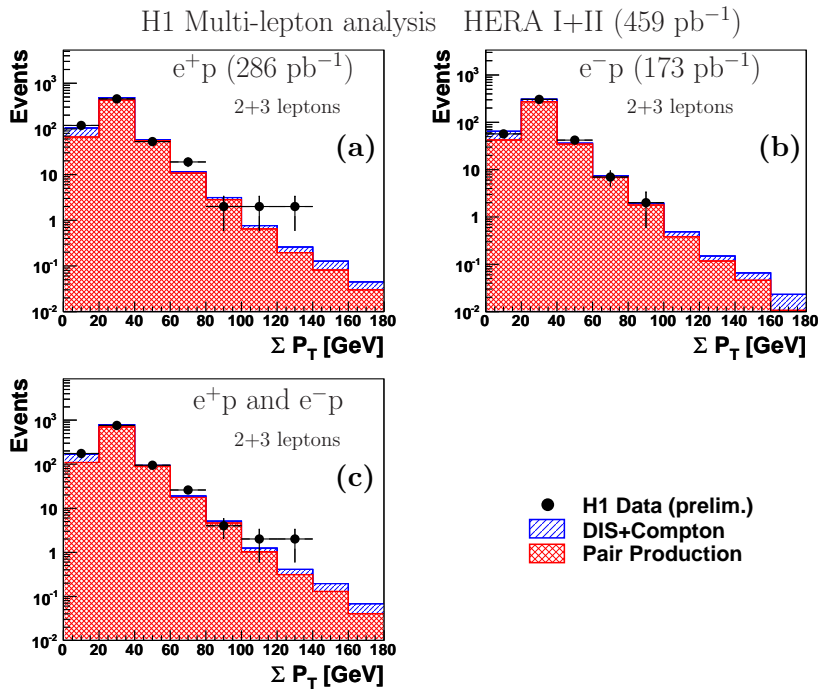}\put(-187,205){{\bf (a)}}\put(-187,80){{\bf (c)}}\put(-27,205){{\bf (b)}}
\end{center}
  \vspace*{-0.3cm}
\caption{Distribution of the scalar sum of the transverse momenta of leptons compared to expectations separately for events recorded by H1 in $e^+p$ (a) and $e^{-}p$ (b) collisions and for all H1 data (c).}
\label{fig:SumEt_All_lep}
\end{figure}

The analysis of di-electron ($2e$) and tri-electron ($3e$) topologies is also carried out by ZEUS using $478$~pb$^{-1}$ of data~\cite{ZEUS_mlep}. 
Two data events with an electron pair invariant mass above $100$~GeV are observed in each $2e$ and $3e$ channel.
These observations are in good agreement with the corresponding SM expectations of $1.9 \pm 0.2$ and $1.0 \pm 0.1$ in the $2e$ and $3e$ channels, respectively.

\begin{figure}[htbp] 
  \begin{center}
    \includegraphics[width=.45\textwidth]{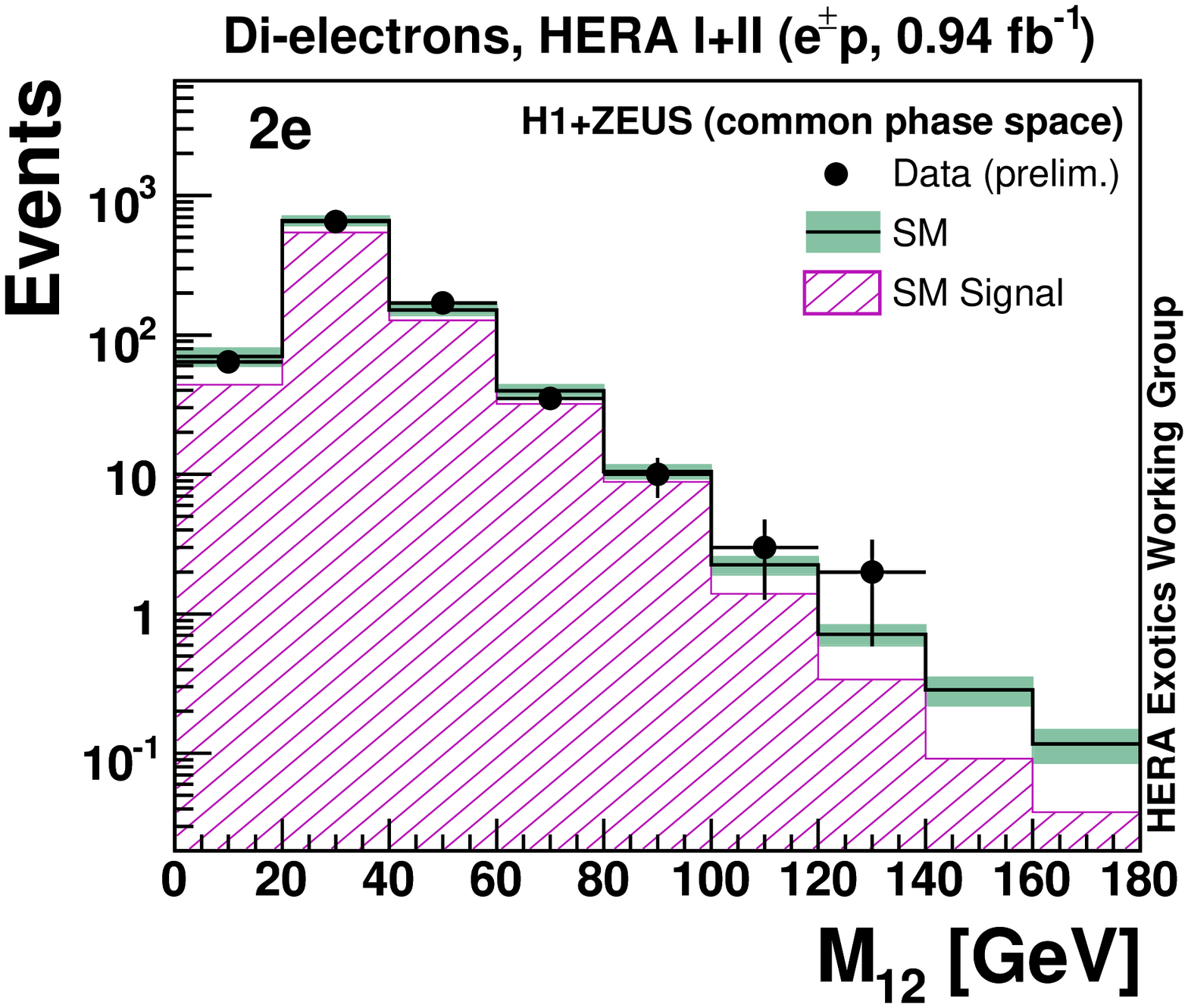}\put(-35,95){{\bf (a)}}
    \includegraphics[width=.45\textwidth]{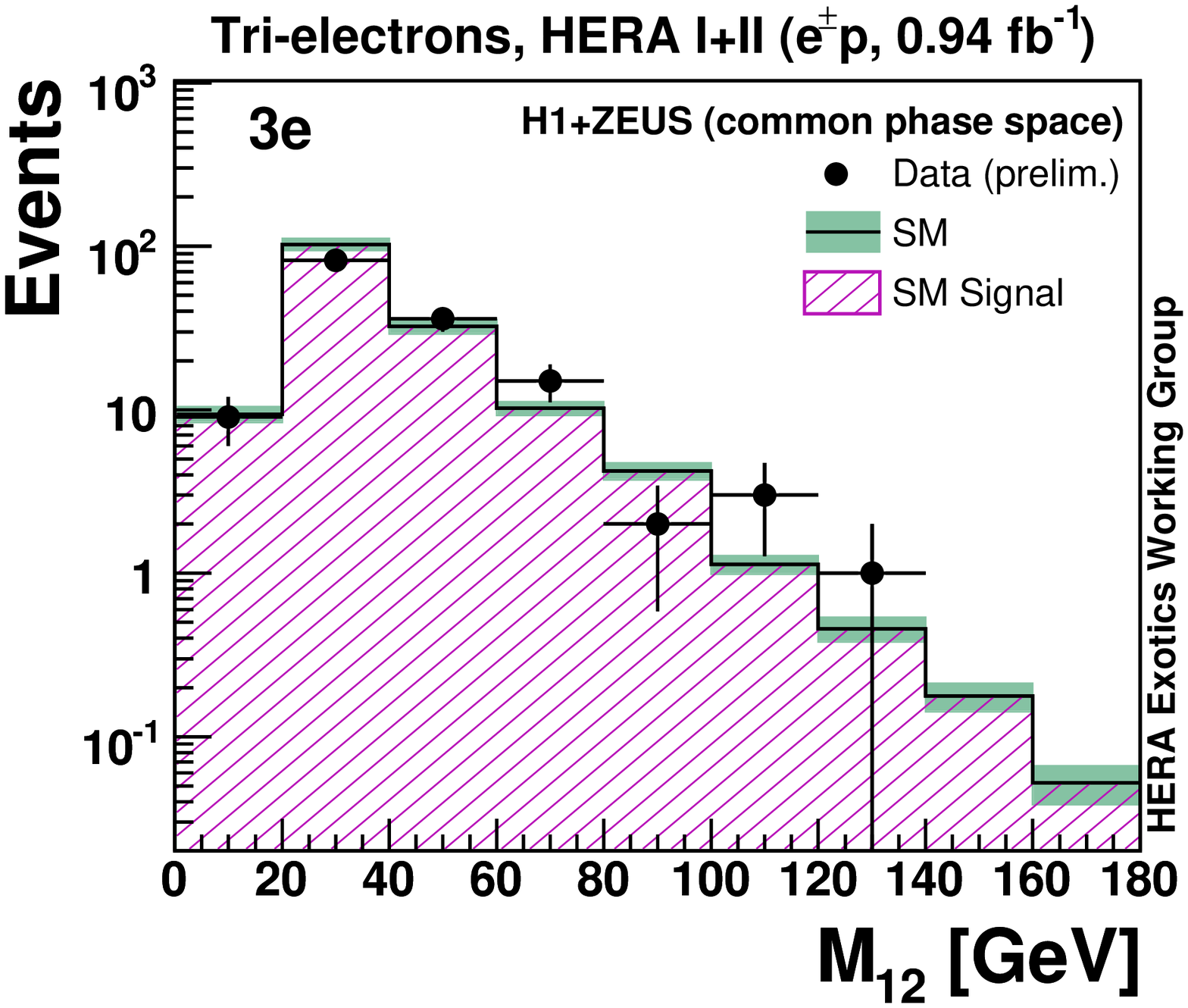}\put(-35,95){{\bf (b)}}
  \end{center}
  \vspace*{-0.5cm}
  \caption{Distribution of the invariant mass $M_{12}$ of the two highest $P_T$ electrons of di-electron~(a) and tri-electron~(b) events observed by H1 and ZEUS in $e^{\pm}p$ data. 
  The points correspond to the observed data events and the open histogram to the SM expectation. The total error on the SM expectation is given by the shaded band. The component of the SM expectation arising from lepton pair production is given by the hatched histogram.}
\label{fig:ML_H1ZEUS}  
\end{figure}

Analyses of the $2e$ and $3e$ topologies from the H1 and ZEUS experiments have been combined in a common phase space~\cite{H1ZEUS_mlep}. The total integrated luminosity amounts to $0.94$~fb$^{-1}$.
The measured event yields of di-electron and tri-electron events are in good agreement with the SM predictions.
The distribution of the invariant mass $M_{12}$ of the two highest $P_T$ electrons in $2e$ and $3e$ channels is presented in figure~\ref{fig:ML_H1ZEUS}.
In the $2e$ ($3e$) channel, $5$ ($4$) events with an invariant mass $M_{12} > 100$~GeV are observed compared to a SM expectation of $3.4 \pm 0.4$ ($1.8 \pm 0.2$).
Combining the two channels, six events are observed with $\sum P_T > 100$~GeV, compared to a SM  expectation of $3.0 \pm 0.3$.
Five of those events  are observed in $e^{+}p$ collisions where the SM expectation is of $1.8 \pm 0.2$, whereas one event is observed in $e^{-}p$ data for a SM prediction of $1.2 \pm 0.1$.

\section{A general search for new phenomena}

\begin{figure}
  \begin{center}
    \includegraphics[width=.6\textwidth,angle=-90]{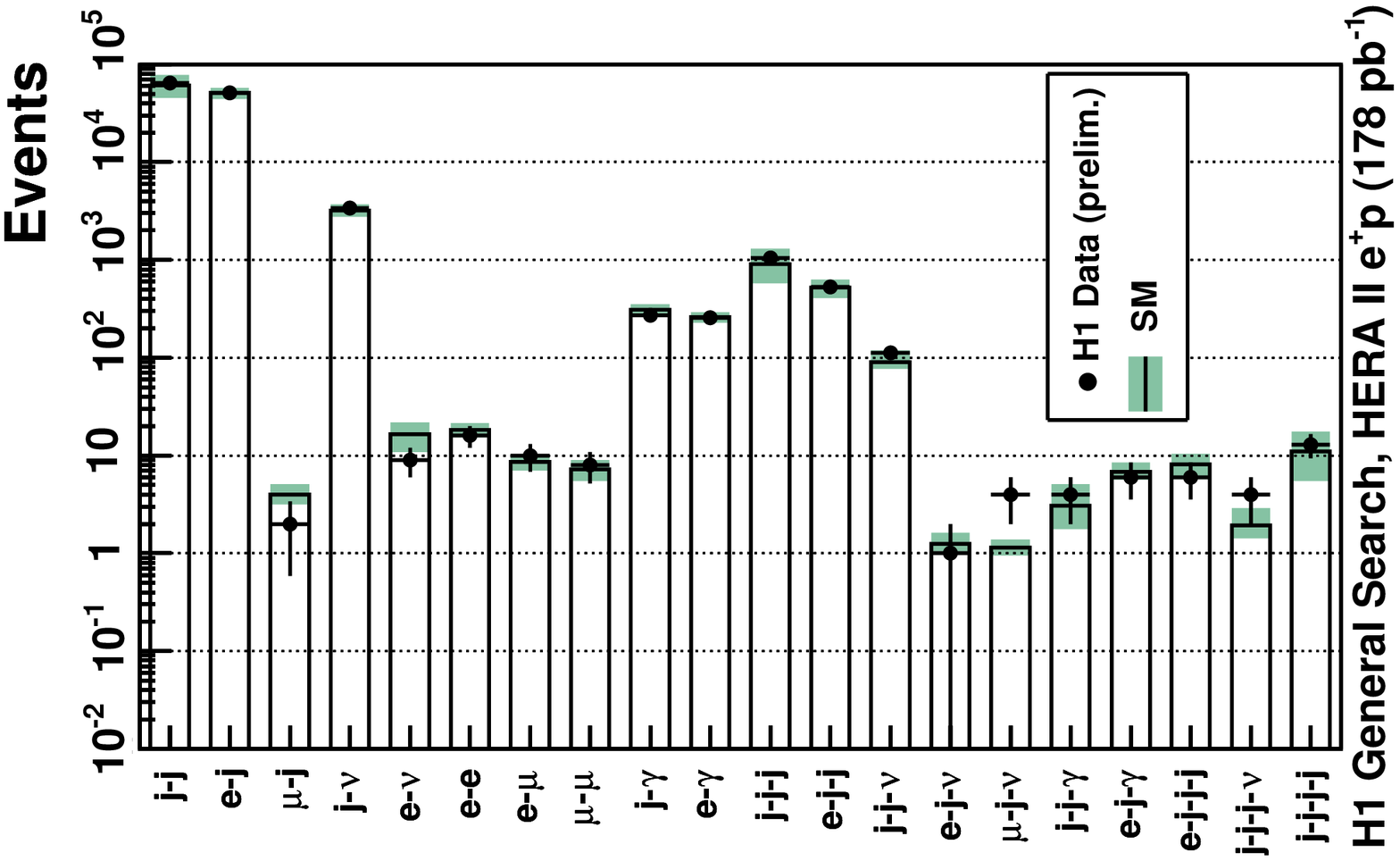}\put(-35,-200){{\bf (a)}}
    \includegraphics[width=.6\textwidth,angle=-90]{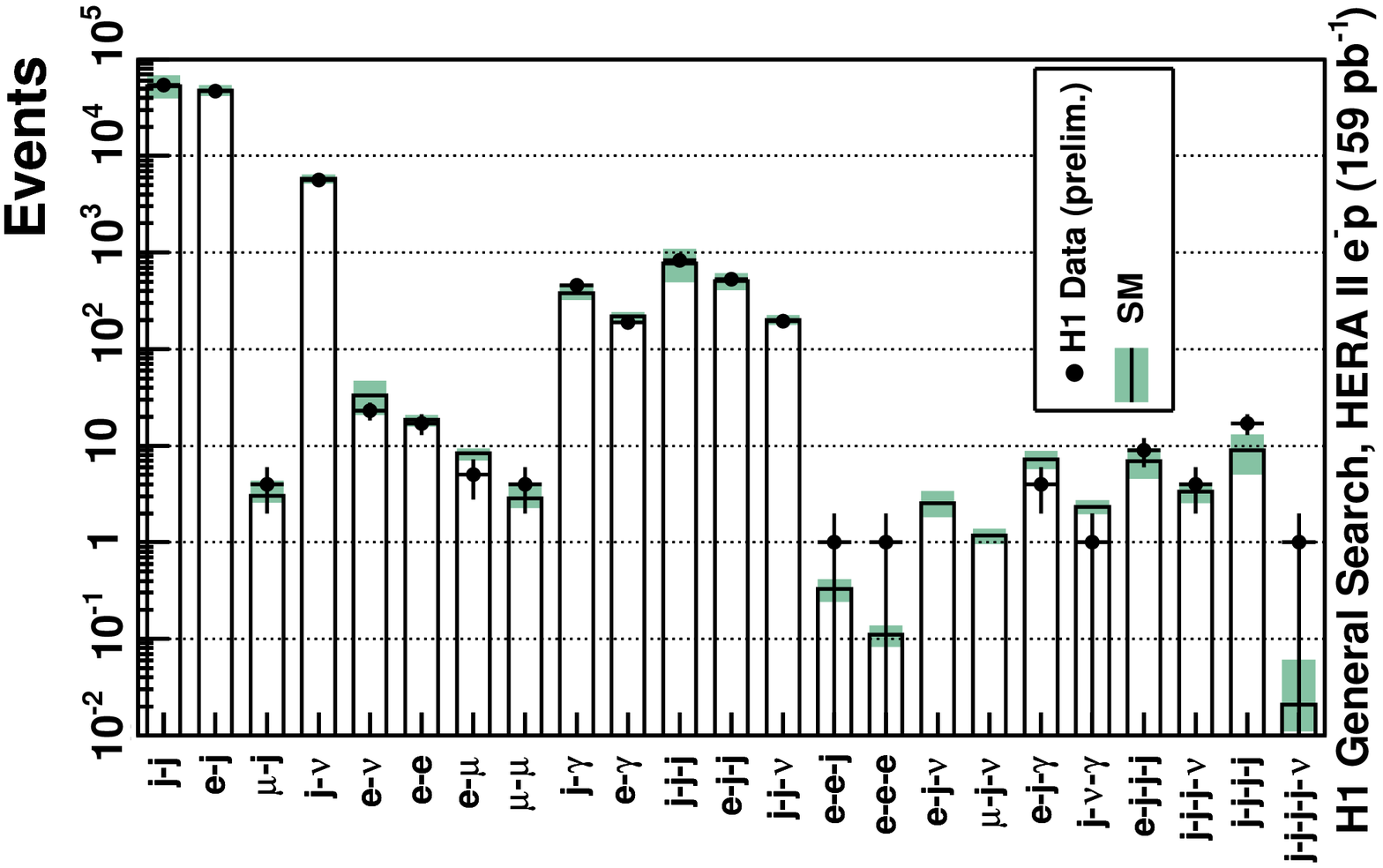}\put(-35,-200){{\bf (b)}}
  \end{center}
  \caption{The data and the SM expectation in event classes investigated by the H1 general search. Only channels with observed data events or a SM expectation greater than one event are displayed. The results are presented separately for $e^+p$ (a) and $e^-p$ (b) collision modes.}
\label{fig:GS}  
\end{figure}

A broad range signature based search has been developed by the H1 Collaboration on HERA~I data~\cite{Aktas:2004pz}. 
All final states containing at least two objects ($e$, $\mu$, $j$, $\gamma$, $\nu$) with 
$P_T >$~$20$~GeV in the polar angle range  $10^\circ < \theta < 140^\circ$ are now also investigated in all HERA~II data~\cite{H1_GS}.
The observed and predicted event yields in each channel are presented in figure~\ref{fig:GS}(a) and (b) for $e^+p$ and $e^-p$ collisions, respectively.
The good agreement observed between data and SM prediction demonstrates the good understanding of the detector and of the contributions of the SM backgrounds. 

A systematic scan of the distributions of the scalar sum of transverse momenta $\sum P_T$ and of the invariant mass $M_{all}$ of all objects is performed in each channel to look for regions of largest deviations from the SM. 
In order to quantify the level of agreement between 
the data and the SM expectation and to identify regions of possible 
deviations, the search algorithm developed in reference~\cite{Aktas:2004pz} is used.
All possible regions in the histograms of $\sum P_T$ and $M_{all}$ distributions are considered. 
A statistical estimator $p$ is defined to judge which region is of 
most interest. This estimator is derived from the convolution of the
Poisson probability density function (pdf) to account for statistical 
errors with a Gaussian pdf to include the effect of 
non negligible systematic uncertainties~\cite{Aktas:2004pz}. 
The value of $p$ gives an estimate of the probability of a fluctuation of the SM expectation upwards (downwards) to at least (at most) the observed number of data events in the region considered.
The region of greatest deviation is the region having the smallest $p$-value, $\pmin$. 
The fact that the deviation could have accured at any point in the distribution is taken into account by calculating the probability $\hat{P}$ to observe a deviation with a $p$-value $\pmin$ at any position in the distribution.
This $\hat{P}$ is a measure of the statistical significance of the deviation observed in the data.
The event class of most interest for a search is the one with the smallest $\hat{P}$ value.

The overall degree of agreement with the SM can further be quantified by 
taking into account the large number of event classes studied in this analysis.
Among all studied classes there is some chance that small $\hat{P}$ values occur. 
This probability can be calculated with MC experiments. 
A MC experiment is defined as a set of hypothetical data histograms following the SM expectation with an integrated luminosity equal to the amount of data recorded.
The complete search algorithm and statistical analysis are
applied to the MC experiments analogously as to the data. 
The expectation for the $\hat{P}$ values observed in the data is then given by the distribution of $\hat{P}^{SM}$ values obtained from all MC experiments.

\begin{figure}[htbp]
\begin{center}
\includegraphics[width=0.42\columnwidth]{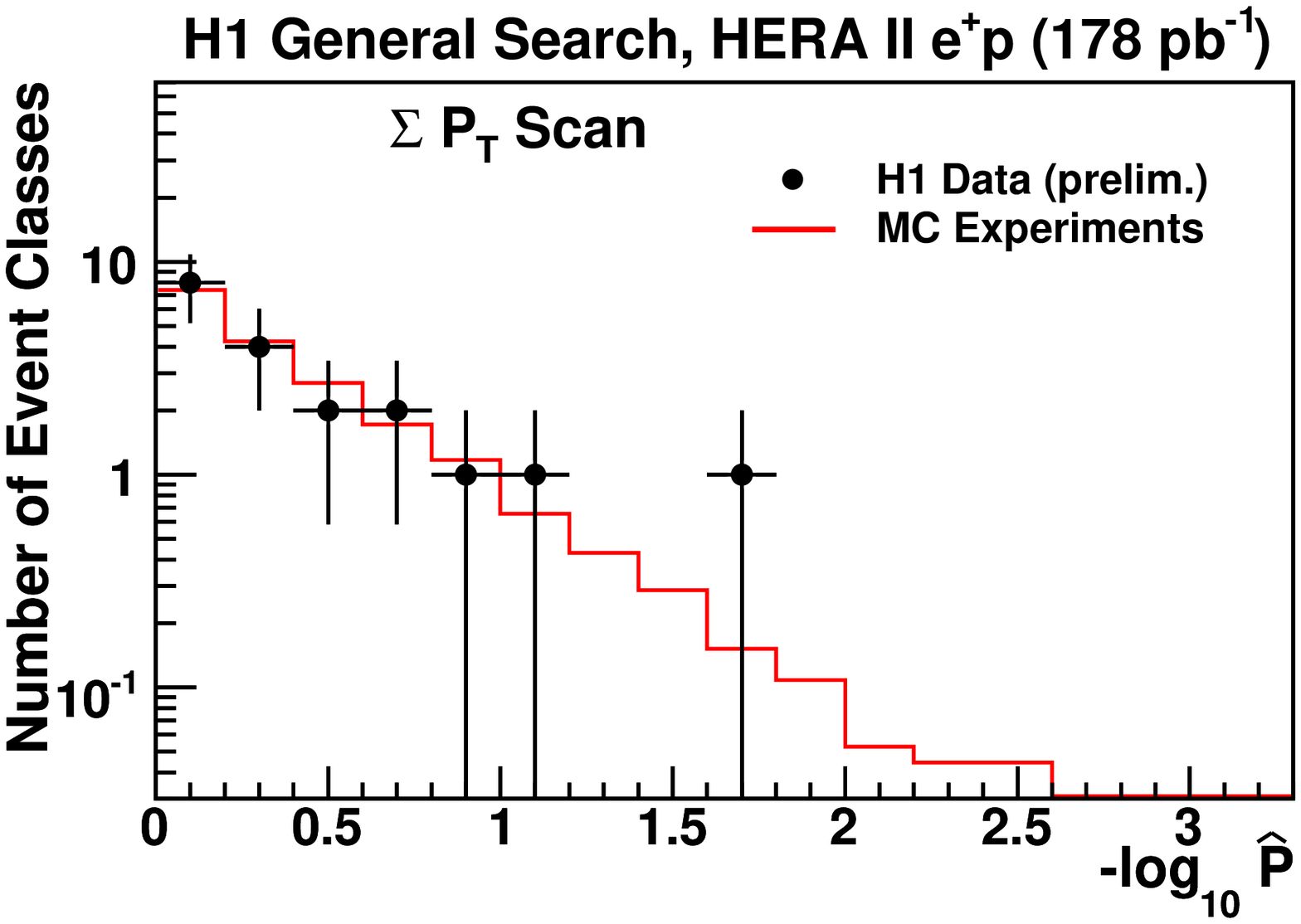}\put(-30,70){{\bf (a)}}
\includegraphics[width=0.42\columnwidth]{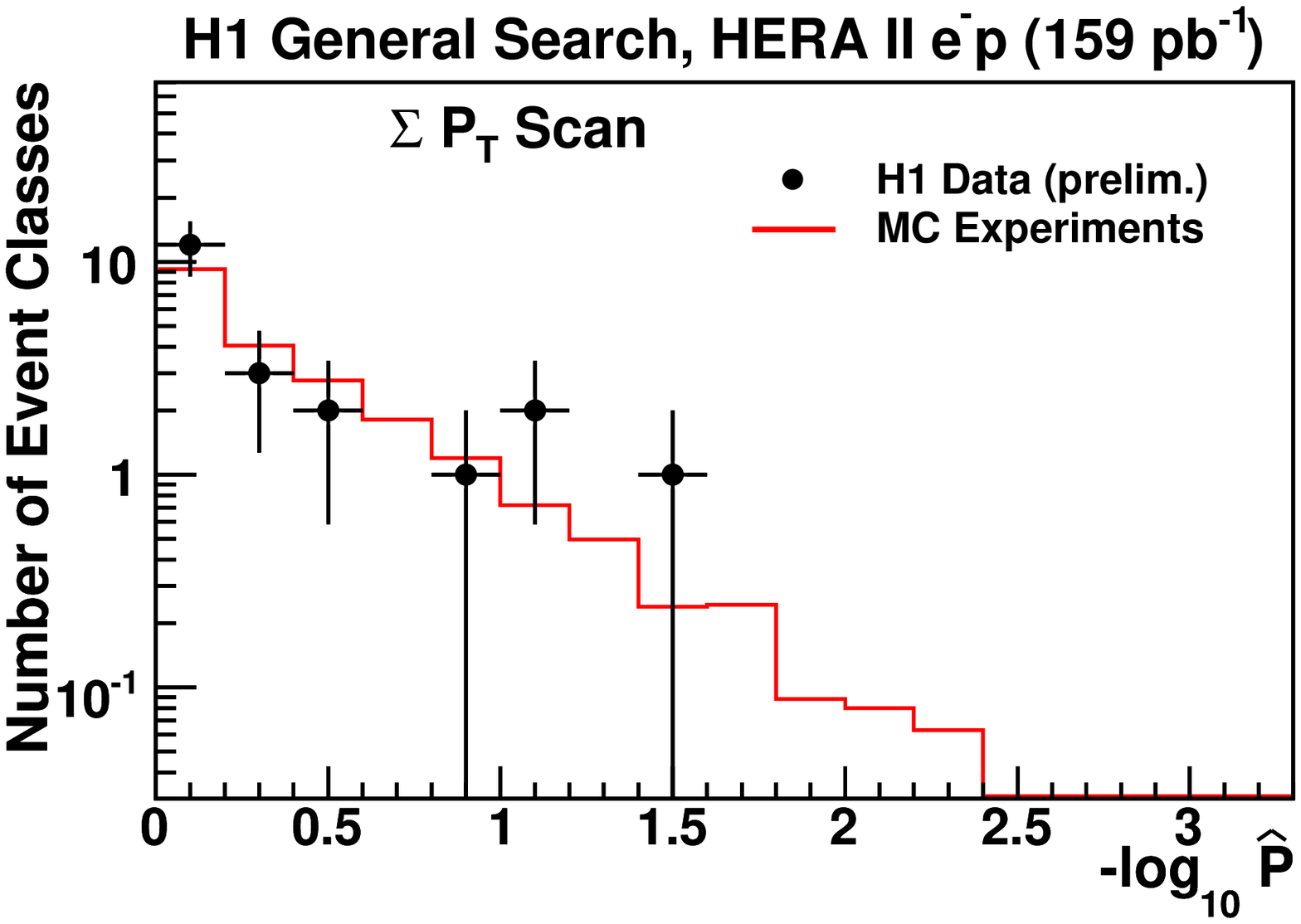}\put(-30,70){{\bf (b)}}
\vspace*{-10pt}
\caption{The $-\log_{10}{\hat{P}}$ values for the data event classes and the 
    expected distribution from MC experiments as derived by investigating
    the $\sum P_T$ distributions in $e^+p$ (a) and $e^-p$ (b) data.}\label{fig:phat}
\end{center}
\end{figure}

The $\hat{P}$ values observed in the real data in all event classes are compared in figure~\ref{fig:phat}  to
the distribution of $\hat{P}^{SM}$ expected from MC experiments. 
The comparison is presented for the scans of the $\sum P_T$ distributions. 
Due to the uncertainties of the SM prediction in the \jjjj~and \jjjjnp~event classes at highest $M_{all}$ and $\sum P_T$ (see reference~\cite{Aktas:2004pz}), where data events are observed, no reliable $\hat{P}$ values can be calculated for these classes. 
These event classes are not considered to search for deviations from the SM in this extreme kinematic domain.
All $\hat{P}$ values range from $0.01$ to $0.99$, corresponding 
to event classes where no significant discrepancy between data and the SM expectation is observed. 
These results are in agreement with the expectation from MC experiments.
The most significant deviation from SM predictions is observed in the \mujnp~event class in $e^+p$ collisions with a value of $-\log_{10}{\hat{P}}$ equal to $1.7$. In the previous H1 analysis~\cite{Aktas:2004pz} based on HERA~I data, which is dominated by $e^+p$ collisions, the largest deviation was also found in this event class, with  $-\log_{10}{\hat{P}}=3$.

\section{Conclusions}

The recent results of model independent searches for new physics performed at the HERA $ep$ collider have been presented.
All analyses fully exploit the complete high energy data sample, which amounts to $\sim 0.5$ fb$^{-1}$ per experiment.
No convincing evidence for the existence of new phenomena beyond the Standard Model has been observed.
Among all event topologies investigated, the largest deviation to the SM expectation is observed by the H1 experiment for isolated lepton events in $e^+p$ collisions. After having analysed all data recorded by H1, this deviation corresponds to a $3$ $\sigma$ excess of atypical $W$-like events. 
This deviation is not confirmed by the ZEUS experiment.

\end{document}